\documentclass{aastex}\usepackage{emulateapj5,epsfig} %apj preprint

\def\bv{{\bf v}}

\def\bb{{\hat{\bf b}}}

\def\tbB{{\tilde{\bf B}}}
\def\tbJ{{\tilde{\bf J}}}

\def\bB{{\bf B}}
\def\bkp{{{\bf k}_{\perp}}}
\def\bxi{{\bf\xi}}
\def\bJ{{\bf J}}

\def\hbtheta{\hat{\mbox{\boldmath$\theta$}} }
\def\bkappa{\mbox{\boldmath$\kappa$}}
\def\bxi{\mbox{\boldmath$\xi$}}
\def\bxip{{{\mbox{\boldmath$\xi$}}_{\perp}}}
\def\etap{{{\mbox{\boldmath$\eta$}}_{\perp}}}
\def\hatbz{\hat{b}_{z}}
\def\hatbr{\hat{b}_{r}}
\def\cs2{{c_{s}^{2}}}

\newcommand{\myunit}[1]{\mbox{$\;\mathrm{#1}$}}

\newcommand{\gram}{\myunit{g}}  %gram
\newcommand{\cm}{\myunit{cm}}   %centimeters
\newcommand{\second}{\myunit{s}} %second
\newcommand{\K}{\myunit{K}}     %degrees Kelvin
\newcommand{\GramPerCc}{\gram\cm^{-3}} %mass density
\newcommand{\gauss}{\myunit{G}} %gauss

\newcommand{\MeV}{\myunit{MeV}} %MeV

\newcommand{\Msun}{\mbox{$M_\odot$}}

\newcommand{\hour}{\myunit{hr}} %hour
\newcommand{\yr}{\myunit{yr}}   %years
\newcommand{\km}{\myunit{km}}   %kilometers

\newcommand{\NA}{N_\mathrm{\!A}} %Avogadro no.
\newcommand{\kB}{k_\mathrm{B}}  %Boltzmann's constant

\newcommand{\EF}{E_\mathrm{F}}  %Fermi energy

\newcommand{\ee}[1]{\times 10^{#1}}

\slugcomment{Submitted to Astrophysical Journal}

\shorttitle{Ballooning Instability in Polar Caps}
\shortauthors{Litwin, Brown, \& Rosner}

\begin{document}

\title{Ballooning Instability in Polar Caps of Accreting Neutron Stars}

\author{C. Litwin, Edward F. Brown, and R. Rosner}
\affil{Department of Astronomy \& Astrophysics, The University of Chicago}
\email{litwin@zohar.uchicago.edu}

\begin{abstract}
   We assess the stability of Kruskal-Schwarzschild (magnetic
   Rayleigh-Taylor) type modes for accreted matter on the surface of a
   neutron star confined by a strong ($\gtrsim 10^{12}$ G) magnetic
   field.  Employing the energy principle to analyze the stability of
   short-wavelength ballooning modes, we find that line-tying to the
   neutron star crust stabilizes these modes until the overpressure at
   the top of the neutron star crust exceeds the magnetic pressure by
   a factor $\sim 8(a/h)$, where $a$ and $h$ are respectively the
   lateral extent of the accretion region and the density scale
   height.  The most unstable modes are localized within a density
   scale height above the crust.  We calculate the amount of mass that
   can be accumulated at the polar cap before the onset of
   instability.
\end{abstract}


\section{Introduction}\label{intro}

Despite recent progress on the spreading of accreted matter from a
disk over an \emph{unmagnetized} neutron star
\citep{inogamov.sunyaev:spread,popham.sunyaev:accretion}, there has
been scarcely any effort directed at understanding how the
magnetically funneled accretion stream of an accretion-powered X-ray
pulsar spreads over the neutron star's surface.  The presence of a
strong magnetic field raises the possibility that the accreted matter
is confined to the polar regions of the neutron star.  This funneling
and confinement of the accreted matter has been proposed
\citep{joss80:_helium} as the reason for the apparent stability of
thermonuclear burning on strongly magnetized neutron stars.  The
presence of strong magnetic fields, with $B\gtrsim 10^{12}\gauss$, is
deduced from cyclotron features in their spectrum \citep[for reviews,
see][]{white.nagase.ea:properties, heindl.coburn.ea:rxte} and from
regular torque reversals \citep{bildsten97b} that suggest the
magnetospheric and co-rotation radii are similar.  In neutron stars
with weak magnetic fields (as inferred by the absence of pulsations or
cyclotron features in the persistent emission), unstable burning is
observed as type~I X-ray bursts \citep{lewin95}.  The funneling and
confinement of the accreted matter by the magnetic field is believed
to increase the local accretion rate to near- or super-Eddington
magnitudes, for which the burning is stable to temperature
perturbations \citep[see][for a review]{bildsten:thermonuclear}.

Confinement of the accreted matter requires its overpressure be
balanced by the magnetic field.  Magnetostatic equilibria of this type
were discussed by \citet{hameury83} and \citet{brown98}.  In these
papers, however, the stability of such equilibria was not addressed. 
This question is considered here.

If the magnetic field has an approximately potential dipole shape (as
can be expected as long as the accreted plasma overpressure is low
compared to the magnetic pressure), the force of gravity has a
nonvanishing component perpendicular to the magnetic field.  Such a
configuration might be expected to be susceptible to
Kruskal-Schwarzschild (``magnetic Rayleigh-Taylor'') interchange
instabilities \citep{kruskal54}.  Indeed, it is easy to show that for
characteristic parameters of polar cap plasmas the stabilizing effect
of the magnetic curvature is small compared to the gravitational
drive.  Nevertheless, the line-tying in the neutron star crust
eliminates interchange modes.  Instead, ballooning modes can be
unstable if the ratio of plasma overpressure to magnetic pressure
$\Delta\beta$ is sufficiently large.  This instability has been a
subject of numerous investigations in the context of laboratory
\citep[e.g.,][]{ohkawa61,coppi66,connor79,freidberg87}, solar
\citep[e.g.,][]{hood86,debruyne89,strauss94}, magnetospheric
\citep[e.g.,][]{mcnutt87,hameiri91} and astrophysical
\citep[e.g.,][]{parker66,shibata89} plasmas.  In the present context
we find that this instability occurs when $\Delta\beta$ exceeds $a/h$
where $a$ is the polar cap radius and $h$ is the density scale height
near the neutron star crust.  For such a large overpressure the
equilibrium magnetic field is strongly distorted away from the
potential field.

The structure of this paper is as follows.  We begin by reviewing
(\S~\ref{polar-caps}) the conditions at the polar cap of an accreting
neutron star.  We then formulate, in section~\ref{model}, the
mathematical model which serves as the framework of our linear
stability analysis.  In section~\ref{energy} we discuss the energy
principle and derive a simple intuitive expression for the
magnetohydrodynamic (MHD) potential energy, relevant for plasmas in the 
neutron star polar caps.  We then use this expression to derive a
criterion for stability of short-wavelength (in the direction
transverse to the magnetic field) ballooning modes and discuss the
stability of a plasma confined by an approximately potential, dipole
magnetic field (\S~\ref{ballooning}).  We then derive an approximate
analytic expression for an equilibrium in which a large overpressure
distorts the field away from the potential field
(\S~\ref{equilibrium}).  Using this equilibrium we find the marginal
stability criterion (\S~\ref{evaluation}), from which we compute the
amount of mass that must be accreted in order to drive the instability
(\S~\ref{mass}).  We conclude with the discussion of our results in
Section~\ref{discussion}.

\section{The structure of polar caps}\label{polar-caps}

We are concerned with the stability of accreted matter on the polar cap
of a strongly magnetized neutron star.  Figure~\ref{fig:schematic}
provides a schematic of the situation.  The collimated infalling matter
approaches the neutron star at roughly free-fall velocity
$v_{\it ff}\sim c$.  At the near- or super-Eddington local accretion
rates, some form of radiative shock is likely to form
\citep{basko.sunyaev:limiting}.  The region we are interested in is far
below where the infalling material decelerates and becomes incorporated
into the neutron star atmosphere.  The continual deposition of matter
into the upper atmosphere pushes the underlying material deeper, and the
downward flow velocity is $v\approx \dot{m}/\rho\ll c_s$, where
$\dot{m}$ is the local accretion rate per unit area.  The atmosphere is,
to a high accuracy, in hydrostatic equilibrium.  On time scales for
matter to flow through the thickness of the atmosphere, the field lines
co-move with the fluid \citep{brown98}.  Under these conditions we are
justified in describing the structure of the accretion region with a
magnetostatic approximation.

The polar cap is characterized by two length scales: its radius and
the density scale height.  Various estimates of the polar cap radius
$a$ \citep{lamb73:_x,elsner77:_accret,arons_lea80} suggest that $a\sim
10^5\cm$, and we shall adopt this value throughout the remainder of
this paper.  The density scale height $h$ is much smaller than this
value of $a$ (see below).  With these approximations, the mass and
radius of the neutron star enter only through the surface gravity $g$,
for which we use $g=1.86\ee{14}\cm\second^{-2}$, which is the
Newtonian value for a canonical neutron star of mass $M=1.4\Msun$ and
a radius $R=10\km$.

\smallskip\epsfig{file=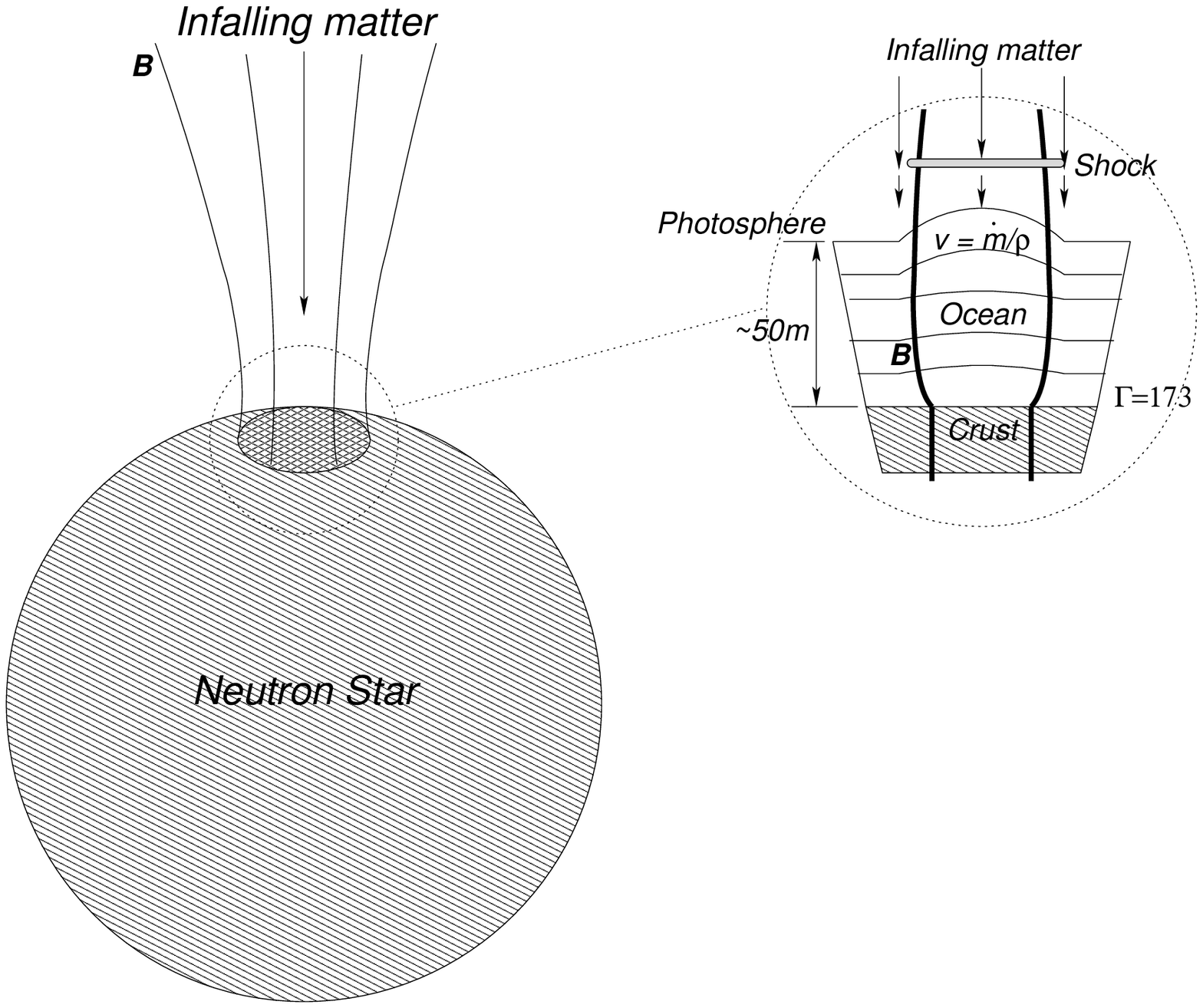,width=8.5cm}
\figcaption{\label{fig:schematic} Overview of the accretion region.}
\medskip

Numerical solutions \citep{hameury83,brown98} of the magnetostatic
equations show that the greatest distortion of the magnetic field
occurs near the boundaries where the field lines are assumed to be
tied.  In the outermost layers of the neutron star, such a condition
occurs at the neutron star crust where the ions form a Coulomb
lattice.  It is the rigidity of the crust that supports the
superincumbent mass of the accreted matter in the polar cap and
anchors the magnetic field lines.  The electrostatic coupling is
parameterized by
\begin{eqnarray}\label{eq:Gamma}
\Gamma  &=&       \frac{Z^2 e^2}{\kB T} \left(\frac{4\pi
                  n_\mathrm{N}}{3}\right)^{1/3} \nonumber \\
        &\approx& 1.1 \left(\frac{Z^2}{A^{1/3}}\right)
                  \left(\frac{\rho}{10^8\GramPerCc}\right)^{1/3}
                  \frac{10^8\K}{T}
\end{eqnarray}
where $k_{B}$ is the Boltzmann constant, $\rho$ is the density, and
$T$, $n_\mathrm{N}$, $Ze$, and $A$ are, respectively, the temperature,
the number density, the charge, and the atomic number of nuclei.  For
$\Gamma\gtrsim 1$, the ions form a liquid.  For sufficiently large
$\Gamma$, the ions form a crystalline lattice.  The composition of the
deep ocean and crust is unknown \citep[see][]{schatz99}; for
definiteness we take the ocean and crust to be composed of pure iron,
and assume that solidification occurs for $\Gamma > 173$, in
accordance with recent Monte-Carlo simulations \citep{farouki93}.  At
this location, the crust of the neutron star can support a shear
stress and therefore anchor the field lines.  In the absence of a full
calculation of the crust and magnetic field, we shall treat the crust
[similarly as in previous works \citep{hameury83,brown98}] as a rigid,
perfectly conducting surface to which the field lines are anchored
(see, however, \S~\ref{discussion}).

As evident in equation~(\ref{eq:Gamma}), the location of the crust is
determined by the temperature of the ocean.  We assume that the burning
of accreted hydrogen and helium is stable (this assumption can be
checked \emph{a posteriori} by finding the depth at which matter
spreads).  Stablity of the H/He burning requires that the accretion rate
be locally super-Eddington \citep[see][]{bildsten:thermonuclear}; for
the remainder of this work we shall use the local Eddington rate as our
fiducial value.  In this case, the temperature in the ocean is set by
the nuclear flux released from burning of the accreted fuel to iron peak
elements (about $8\MeV$ per nucleon) and is $\gtrsim5\ee{8}\K$
\citep{brown98,schatz99}.

For the density and temperature of interest, the electrons are
degenerate, i.e., $\kB T < E_F - m_e c^2$, where $E_{F} = m_e c^2
[{1+(3\pi^2 n_e)^{2/3} (\hbar/m_e c)^2}]^{1/2}$ is the electron Fermi
energy, and $m_e$ and $n_e$ are the electron mass and density,
respectively.  The principal contribution to pressure $p$ is the
electron pressure $p_{e}$ and therefore to a good approximation
$p\approx p_{e}\approx n_e\EF/4\propto \rho^{\gamma}$ where $\gamma
=4/3$ and the proportionality constant is only a function of the ion
charge-to-mass ratio $Z/A$.  For a plasma of uniform composition this
constant is independent of both time and space.  We shall use this
approximation in the next section.

Using this approximate equation of state, we write the density scale
height $h=\gamma p/\rho g$ at the top of the crust in terms of $\Gamma$,
\begin{eqnarray}\label{scaleheight}
h &\approx& \gamma\left(\frac{9\pi}{256}\right)^{1/3} \frac{1}{Z^{2/3}A}
        \frac{\NA\kB T}{g}\frac{\Gamma}{\alpha_\mathrm{F}}\nonumber\\
   &=&  6.90\ee{3}\cm\nonumber\\
        &&\times
	\left(\frac{26}{Z}\right)^{2/3}
        \left(\frac{56}{A}\right) \left(\frac{T}{5\ee{8}\K}\right)
        \left(\frac{\Gamma}{173}\right),
\end{eqnarray}
where $\gamma =4/3$, $\NA=6.02\ee{23}\gram^{-1}$ is Avogadro's
constant and $\alpha_\mathrm{F}=e^2/\hbar c$.

%\vspace{12pt} 
\section{Model}\label{model}~{ We model the polar cap
as the region of the neutron star atmosphere and ocean that is bounded
by the same magnetic flux surface within which accretion occurs.  This
region is characterized by a somewhat higher pressure than in the
surrounding matter, due to the weight of the accreted material.  The
plasma in the polar cap and in the surrounding region will be treated
as a perfectly conducting fluid of uniform composition, bounded by a
perfectly conducting, rigid surface (the top of the crust) and immersed
in a magnetic field.  The magnetic field lines are assumed tied to the
crust; the intersection point of the field line with the top of the
crust will be referred to as the footpoint.  Since the density scale
height of the plasma in the polar cap is much smaller than the stellar
radius we shall consider the field lines as semi-infinite and having
only one footpoint.  In the equilibrium, the plasma and the immersing
magnetic field are assumed to be axisymmetric with the axis of symmetry
coinciding with the magnetic axis; the magnetic field is assumed to be
untwisted.}

We describe the plasma by the single-fluid MHD theory.  In the
presence of the gravitational potential $U$ and magnetic field $\bB$
the momentum balance equation for a plasma of density $\rho$ and
pressure $p$, moving with velocity $\bv$, is
\begin{equation}\label{eq1}
\rho\frac{d\bv}{dt}=\frac{1}{c}\bJ\times\bB - \nabla p -\rho\nabla U.
\end{equation}
While Ohm's law is modified by gravity, the induction equation is
unaffected (cf.\ \citet{bernstein58}), so that
\begin{equation}\label{eq2}
\frac{\partial\bB}{\partial t}=\nabla\times\bv\times\bB.\end{equation}

\noindent The above equations are supplemented by Amp\`ere's law,

\begin{equation}\label{ampere}
\bJ =\frac{c}{4\pi}\nabla\times\bB,
\end{equation}

\noindent the continuity equation,

\begin{equation}\label{continuity}
\frac{\partial\rho}{dt}+\nabla\cdot (\rho\bv )=0,
\end{equation}

\noindent and an equation of state which we assume to be adiabatic:

\begin{equation}\label{eos}
\frac{d}{dt}p\rho^{-\gamma}=0.
\end{equation}

As mentioned in \S\ref{intro}, the primary contribution to the plasma
pressure in the region of interest is due to degenerate, relativistic
electrons ($\gamma =4/3$).  Therefore, in region of uniform composition,
$p\rho^{-\gamma}$ is, to a high accuracy, constant in space.  For the
assumptions in \S~\ref{polar-caps}, $\epsilon\equiv d\ln\rho
p^{-1/\gamma}/d\ln p \approx 3/8 p_\mathrm{ion}/p_\mathrm{e} = (3/2Z)
(\kB T/\EF) \approx 4\times 10^{-4}$.  Thus

\begin{equation}\label{equilibrium_eos}
p\rho^{-\gamma}=\mathrm{const.},
\end{equation}

\noindent to a very good approximation.

The force balance in a static equilibrium is given by

\begin{equation}\label{eq3}
\frac{1}{c}\bJ\times\bB = \nabla p +\rho\nabla U,
\end{equation}

\noindent from which it follows that

\begin{equation}\label{eq4}
\nabla p +\rho\nabla U = \rho\nabla F,
\end{equation}

\noindent where the function $F$ is a function constant on the field lines

\begin{equation}\label{eq5}
    \bB\cdot\nabla F =0.
\end{equation}

\noindent By setting the gravitational potential to zero at the top of
the crust, $F$ is determined by the sound speed $c_{s}=\sqrt{\gamma
p/\rho}$ there, $F=c_{s0}^{2}/(\gamma -1)$.

An untwisted, axisymmetric equilibrium magnetic field can be
represented in the form

\begin{equation}\label{eq5a}
\bB = \nabla\psi\times\nabla\theta
\end{equation}

\noindent where $\psi$ is the flux function and $\theta$ is the toroidal
angle.  It is straightforward to show with the aid of Amp\`ere law that
in such a geometry the current is perpendicular to the magnetic field:
$\bJ\cdot\bB =0$.  Eq.~(\ref{eq5}) implies that $F$ is a function of
the flux function only: $F=F(\psi )$.

Upon introducing cylindrical coordinates ($r, \theta, z)$ with the
$z$-direction along the axis of symmetry, the normal (i.e., parallel
to $\nabla\psi$) component of Eq.~(\ref{eq4}) yields the
Grad-Shafranov-like equation

\begin{equation}\label{eq5b}
r\frac{\partial}{\partial r}\frac{1}{r}\frac{\partial\psi}{\partial r} +\frac{\partial^{2}\psi}{\partial
z^{2}}=-4\pi r^{2}\rho (\psi ,z)\frac{d F}{d\psi}.
\end{equation}

\noindent with $\rho$ being determined by equation (\ref{eq4}).

In the limit $a\ll R$ the curvature of the stellar surface can be
neglected on the scale of the accretion region.  In this
approximation $U=g z$, so that

\begin{equation}\label{density}
\rho (\psi,z)=\rho_{0}(\psi)\left[ 1-\alpha
\frac{z}{h(\psi )}\right]^{{1}/{\alpha}} 
\end{equation}
where $z=0$ corresponds to the top of the crust, $\alpha\equiv\gamma -1$,
$\rho_{0}$ is the plasma density at $z=0$, $h\equiv -(d\ln\rho
/dz)^{-1}|_{z=0}=\alpha F/g$ is the density scale height at the
footpoints and $g$ is the gravitational acceleration.  Note that
$\rho_{0}$ and $h$ are functions of $\psi$.

\section{Energy Principle}\label{energy}

We now address the question of the linear stability of the polar cap
plasma in the approximation described in the previous section.  We
base our analysis on the MHD energy principle (Bernstein et al.\
1958).  In the present section we discuss the energy principle as
applicable to the polar cap plasmas, generalizing the ``intuitive''
form of the energy principle \citep{furth65,greene68} to include the
effects of a strong gravitational field.  In our discussion we follow
the approach of \citet{freidberg87}.

We start by considering the linearized momentum balance equation for a
static equilibrium

\begin{equation}
\rho\frac{\partial^2\bxi}{\partial t^2}={\bf F}
(\bxi)\equiv\frac{1}{c}\tbJ\times\bB + \frac{1}{c}\bJ\times\tbB -
\nabla\tilde{p} - \tilde{\rho}\nabla U
\end{equation}

\noindent where $\bxi$ is the plasma displacement and the gravitational
potential is unperturbed.  Here and in the following the tilde denotes
perturbed quantities while symbols without it stand for the equilibrium
quantities.  Perturbed current $\tbJ$, magnetic field $\tbB$, density
$\tilde{\rho}$, and pressure $\tilde{p}$ are respectively given by
Amp\`ere's law, the induction equation, the continuity equation, and the
equation of state:

\begin{eqnarray}
\tbJ =\frac{c}{4\pi}\nabla\times\tbB\\
\tbB =\nabla\times (\bxi\times\bB )\\
{\tilde \rho}=-\nabla\cdot (\rho\bxi )\\
{\tilde p}=-\gamma p\nabla\cdot\bxi -\bxi\cdot\nabla p.
\end{eqnarray}

\noindent We shall assume that the plasma boundary (the crust) is rigid
and perfectly conducting, so that $\bxi$ and $\tbB$ vanish there.

The second-order potential energy is given by (Bernstein et al.\ 1958)

\begin{eqnarray}
\delta W=-\frac{1}{2}\int d\tau\bxi\cdot{\bf F}(\bxi )\nonumber\\
=-\frac{1}{2}\int d\tau \bxi\cdot\left\{\frac{1}{c}\tbJ\times\bB
+\frac{1}{c}\bJ\times\tbB\nonumber\right.\\ 
\left. \phantom{\frac{1}{c}}+ \nabla (\gamma
p\nabla\cdot\bxi
+\bxi\cdot\nabla p) +\nabla\cdot (\rho\bxi )\nabla U\right\}
\end{eqnarray}

\noindent where $d\tau$ denotes integration over the plasma volume. 
We now transform the above expression for the potential energy into a
form that allows a physical interpretation.  This is analogous to what
\citet{furth65} and \citet{greene68} have done in the absence of
gravitational field.  In order to derive it, we follow the standard
approach [e.g., \citet{freidberg87}], with appropriate modifications for
the presence of a strong gravitational field.

Integrating by parts and dropping surface terms (which vanish because
of the assumed boundary conditions), and exploiting the equilibrium
relation given by Eq.~(\ref{eq4}), one obtains, after some
straightforward algebra,

\begin{eqnarray}\label{energy-1}
\displaystyle &\delta W =\frac{1}{2}\int d\tau
\left\{\gamma p(\nabla\cdot\bxi -2\bkappa_{g}\cdot\bxi )^{2} 
+\epsilon p(\xi\cdot\nabla\ln p)^{2}+\frac{\tbB^{2}}{4\pi}
\right.\nonumber\\
&\left. -\bxi\cdot\left[\frac{1}{c}\bJ\times\tbB +
2\bkappa_{g}(\rho\bxi\cdot\nabla F)+\nabla (\rho\bxi\cdot\nabla F)\right]\right\}.
\end{eqnarray}

\noindent where $\bkappa_{g}\equiv(\rho /\gamma p)\nabla U/2$.  The
the second term on the RHS, proportional to the Schwarzschild
discriminant, can be neglected for highly degenerate plasmas in strong
magnetic fields because of the smallness of parameter $\epsilon
=d\ln\rho p^{-1/\gamma}/d\ln p$.  We shall omit this term in the
subsequent analysis (see, however, discussion of finite $\epsilon$
effects in \S\ref{discussion}).

The last two terms in the square bracket on the RHS of
Eq.~(\ref{energy-1}) can be expanded and recombined to yield an
alternative form of the energy integral

\begin{eqnarray}\label{energy-2}
\delta W=\frac{1}{2}\int d\tau
\left\{\frac{\tbB^{2}}{4\pi}+\gamma p(\nabla\cdot\bxi-
2\bkappa_{g}\cdot\bxi)^{2}\right.\nonumber\\
\left. - \bxi\cdot \left[\frac{1}{c}\bJ\times\tbB 
+\frac{\rho^{2}}{\gamma p}
(\bxi\cdot\nabla F)\rho\nabla F+\rho\nabla (\bxi\cdot\nabla F)\right]\right\}.
\end{eqnarray}

\noindent Since $\bB\cdot\nabla F =0$ and

\begin{equation}
\frac{1}{c}\bB\cdot\bJ\times\tbB =-\rho\bB\cdot\nabla (\bxi\cdot\nabla F),
\end{equation}

\noindent it is clear that the parallel (to $\bB$) component of the
expression in the square bracket in Eq.~(\ref{energy-2}), and hence
also in Eq.~(\ref{energy-1}), vanishes.  Thus the energy integral can
be expressed in the form

\begin{eqnarray}\label{energy-3}
&\delta W=\frac{1}{2}\int d\tau
\left\{\frac{\tbB^{2}}{4\pi}+\gamma p(\nabla\cdot\bxi-
2\bkappa_{g}\cdot\bxi)^{2}\right.\nonumber\\
&\phantom{\frac{1}{c}}\left.- \bxip\cdot\left[\frac{1}{c}\bJ\times\tbB +
2(\rho\bxip\cdot\nabla F)\bkappa_{g}%\nonumber\right.\\ \left.
+\nabla (\rho\bxip\cdot\nabla
F)\right]\right\}.
\end{eqnarray}

\noindent Next, splitting up $\tbB$ into transverse and parallel
components, $\tbB =\tbB_{\perp}+\tilde{B}_{\parallel}\bb$ ($\bb\equiv\bB
/B$), we find

\begin{eqnarray}
\bxip\cdot\bJ\times\tbB =J_{\parallel}\bxip\cdot\bb\times\tbB +
{\tilde B}_{\parallel} \bxip\cdot\bJ\times\bb\\
\tilde{B}_{\parallel}=-B(\nabla\cdot\bxip +2\bkappa_{c}\cdot\bxip )+
\frac{4\pi\rho}{B}\bxip\cdot\nabla F,
\end{eqnarray}

\noindent where $\bkappa_{c} =\bb\cdot\nabla\bb$ is the field line
curvature.  With these expressions the potential energy can be expressed
in the following, ``intuitive'' form

\begin{eqnarray}\label{intuitive}
\delta W=\frac{1}{2}\int d\tau\left\{\frac{\tbB_{\perp}^{2}}{4\pi} +
\frac{\bB^{2}}{4\pi}(\nabla\cdot\bxip+2\bkappa_{c}\cdot\bxip)^{2}\right.\nonumber\\
+ \gamma p(\nabla\cdot\bxi-2\bkappa_{g}\cdot\bxi)^{2} -
\frac{1}{c}J_{\parallel}\bxip\times\bb\cdot\tbB\nonumber\\
\left.-2\bkappa\cdot\bxip\rho
        (\bxip\cdot\nabla F)\right\}.
\end{eqnarray}

\noindent where $\bkappa\equiv\bkappa_{c} + \bkappa_{g}$.  The first
term is the energy corresponding to field line bending, the second is
the field compression energy, the third is the energy of plasma
compression and buoyancy, the fourth one is responsible for the kink
instability, and the last one drives the curvature and gravitational
instabilities.  The above expression is a generalization of the result
of \citet{furth65} and \citet{greene68} to the case of a degenerate
plasma in a strong gravitational field.  The above expression reduces
to these earlier results if the density scale height is much longer
than the plasma length, i.e., if $U\ll\cs2$.

Thus far we have not taken into account the geometry of polar caps and
the derived energy principle (eq. [\ref{intuitive}]) applies to an
arbitrary configuration of plasmas satisfying the equation of state
(\ref{eos}).  We shall now perform a reduction of the energy 
principle as applicable to the configuration of our interest.

First we note that the parallel displacement $\xi_{\parallel}$ appears
only in the nonnegative compressive term.  Therefore this term can be
minimized, in a standard manner, by considering the perturbation of
the compressive energy due to a perturbation of $\xi_{\parallel}$. 
The minimum is found for the displacement satisfying the 
Euler-Lagrange equation which can be expressed in the form

\begin{equation}
\bB\cdot\nabla\left[p^{1-\frac{1}{\gamma}}\left(\nabla\cdot\bxi -
2\bkappa_{g}\cdot\bxi\right)\right] =0
\end{equation}

Integrating along the field line and imposing the line-tying 
boundary condition ($\xi_{\parallel}[0]=0$) yields

\begin{equation}\label{parallel_displacement}
\xi_{\parallel}=\frac{B}{p^{\frac{1}{\gamma}}}\int_{0}^{l}
\frac{dl'}{B}p^{\frac{1}{\gamma}}\left[ p^{\frac{1}{\gamma}-1}A
+\nabla\cdot\bxip - 2\bkappa_{g}\cdot\bxip\right]
\end{equation}

\noindent where $l$ is the distance along the field line.  The
integration constant $A$ must be determined from the
integrability condition.  Note that $p, \rho\rightarrow 0$ as
$z\rightarrow z_{0}\equiv h/(\gamma -1)$ (cf.  equations
[\ref{density}] and [\ref{equilibrium_eos}]).  Therefore, for the
kinetic energy of parallel motion to be finite, (i.e., $\int
d\tau\rho\xi_{\parallel}^{2}<\infty$), this requires that

\begin{equation}\label{int_const}
A =-\frac{\int_{0}^{l_{0}}
\frac{dl'}{B}p^{\frac{1}{\gamma}}\left(\nabla\cdot\bxip -
2\bkappa_{g}\cdot\bxip\right)}{\int_{0}^{l_{0}}
\frac{dl'}{B}p^{\frac{2}{\gamma}-1}}
\end{equation}

\noindent where $l_{0}\equiv l(z_{0})$.  This implies, in particular,
that $\xi_{\parallel}(z_{0})=0$. Since

\begin{equation}
\nabla\cdot\bxi - 2\bkappa_{g}\cdot\bxi = p^{\frac{1}{\gamma}-1}A
\end{equation}

\noindent and $A\neq 0$, the compressional energy cannot be minimized
to zero.  This is analogous to that which occurs in flux tubes
line-tied at both ends (cf.  Hameiri et al 1991) and is, in fact,
caused by the effective line-tying in the low density region.

Setting $J_{\parallel}=0$, as appropriate for an untwisted
axisymmetric field, the potential energy becomes

\begin{eqnarray}\label{compressive}
\delta W=\frac{1}{2}\int d\tau \left\{\frac{\tbB_{\perp}^{2}}{4\pi} +
\frac{\bB^{2}}{4\pi}(\nabla\cdot\bxip +
2\kappa_{c}\cdot\bxip)^{2}\right.\nonumber\\
+ \gamma p^{\frac{2}{\gamma}-1}\left[\frac{\int_{0}^{l_{0}}
\frac{dl'}{B}p^{\frac{1}{\gamma}}\left(\nabla\cdot\bxip -
2\bkappa_{g}\cdot\bxip\right)}{\int_{0}^{l_{0}}
\frac{dl'}{B}p^{\frac{2}{\gamma}-1}}\right]^{2}\nonumber\\
\left.\phantom{\frac{\tbB_{\perp}^{2}}{4\pi}}-2\rho (\bxip\cdot\nabla
F)\bkappa\cdot\bxip \right\}.
\end{eqnarray}

\section{Ballooning instability}\label{ballooning}

We now limit our attention to the short wavelength (in the direction
perpendicular to the magnetic field) modes.  Following the method of
\citet{freidberg87}, we perform the reduction of the intuitive form of
the energy principle (eq.  [\ref{compressive}]).  

We represent the displacement in the eikonal approximation

\begin{equation}
\bxip =\etap e^{iS} + c.c.
\end{equation}

\noindent where $c.c.$ stands for complex conjugate; $\etap$ is
assumed to be slowly varying while the eikonal $S$ is assumed to vary
rapidly, i.e.,

\begin{equation}
\bkp =\nabla S
\end{equation}

\noindent is so large that $k_{\perp}a\gg 1$, where $a$ is the
equilibrium length scale.

Note that $\bB\cdot\nabla S =0$. Consequently

\begin{equation}
\tbB_{\perp}=e^{iS}[\nabla\times (\etap\times\bB)]_{\perp}+c.c.
\end{equation}

\noindent In the limit of $k_{\perp}\rightarrow\infty$, the dominant
terms in $\delta W$ (Eq.[\ref{compressive}]) are those proportional to
$(\nabla\cdot\bxip)^{2}\approx|\bkp\cdot\etap|^{2}$ (second and third
term on the RHS of eq.  [\ref{compressive}]).  Therefore the
minimization of the potential energy, to lowest order in
$1/k_{\perp}a$, requires that $\bkp\cdot\etap =0$, which implies that
for the zeroth-order displacement

\begin{equation}\label{zeroth}
\etap\approx\frac{X}{B}\bb\times\bkp.
\end{equation}

\noindent where $X$ is a scalar function.  To the next order, the
first-order displacement can be chosen so that the sum of two positive
terms involving $\nabla\cdot\bxip$ (second and third term on the RHS
of eq.  [\ref{compressive}]) is minimized. Since we are interested 
in the limit of $\beta\equiv 8\pi p/B^{2}\gg 1$, we can instead 
minimize only the second of these terms. A simple choice of a 
minimizing displacement is 

\begin{equation}\label{first_order}
\nabla\cdot\bxip = 2\bkappa_{g}\cdot\bxip.
\end{equation}

For this choice, $\xi_{\parallel} =0$ as well as the third term on 
the RHS of eq. (\ref{compressive}) vanish so that

\begin{eqnarray}\label{noncompressive}
    \delta W=\frac{1}{2}\int d\tau
   \left\{\frac{\tbB_{\perp}^{2}}{4\pi} +
\frac{\bB^{2}}{\pi}(\bkappa\cdot\bxip)^{2}\right.\nonumber\\
\left.\phantom{\frac{\tbB_{\perp}^{2}}{4\pi}}
-2\rho (\bxip\cdot\nabla F)(\bkappa\cdot\bxip )\right\}.
\end{eqnarray}

From equation (\ref{zeroth}) it follows that

\begin{equation}
\tbB_{\perp}=e^{iS}(\nabla X\times\bb )_{\perp}+c.c.=(\bb\cdot\nabla
X)\bb\times\bkp+c.c.
\end{equation}

\noindent Consequently, one finds from equation
(\ref{noncompressive}), exploiting the hermiticity of the operator
${\bf F}$ (Bernstein et al.  1958), that

\begin{eqnarray}
    \delta W=\frac{1}{4\pi}\int d\tau \left\{
    k^{2}_{\perp}|\bb\cdot\nabla
    X|^{2}+4(\bkappa\times\bb\cdot\bkp )^{2}|X|^{2}\right.\nonumber\\
    \left.-\frac{8\pi \rho}{B^{2}}
   (\bkappa\times\bb\cdot\bkp ) (\nabla
   F\times\bb\cdot\,\bkp) |X|^{2}\right\}.
\end{eqnarray}

\noindent For an axisymmetric system, $S=m\theta + \hat{S}(\psi ,l)$
where $l$ is the distance along the field line. In the absence of
a toroidal field $B_{\theta}$, $\bB\cdot\nabla S=0$ implies that
$\hat{S}=\hat{S}(\psi )$.  Thus

\begin{equation}
    \bkp =k_{\theta}\hbtheta +k_{n}{\hat{\bf n}}
\end{equation}

\noindent where $k_{\theta}=m/r$ and ${\hat{\bf n}}=\nabla\psi
/|\nabla\psi |=\nabla\psi /rB$ is the unit vector normal to the flux
surface.  Noting that

\begin{eqnarray}
    \bb\times\bkp =-k_{\theta}{\hat{\bf n}}+k_{n}\hbtheta\\
      \bb\times\bkp\cdot\nabla F =-k_{\theta}rB\frac{d F}{d\psi},
\end{eqnarray}

\noindent the potential energy can be put in form

\begin{eqnarray}\label{eq32}
    \delta W=\frac{1}{4\pi}\int d\tau
    \left\{(k_{\theta}^{2}+k_n^2)\left|\frac{\partial X}{\partial l}\right|^{2}
    \right.\nonumber\\\left.
    +4k_{\theta}^{2}\kappa_{n}(\kappa_{n}-\frac{2\pi r\rho 
    F'}{B})|X|^{2}\right\}
\end{eqnarray}

\noindent ($F' ={d F}/{d\psi}$).  Minimization with respect to $k_{n}$
is achieved by setting $k_{n}=0$.

The integrand in Eq.~(\ref{eq32}) involves derivatives of $X$ only
along the magnetic field but not derivatives of $\psi$.  Hence $\psi$
can be treated as a parameter and the potential energy can be
expressed in the form

\begin{eqnarray}\label{separation}
    \delta W=\int d\psi\delta W(\psi ),
\end{eqnarray}

\noindent where

\begin{eqnarray}\label{crit}
   \delta W(\psi )=\frac{m^{2}}{4\pi}\int\frac{dl}{r^{2}B}
    \left\{\left|\frac{\partial X}{\partial l}\right|^{2}
   \right.\nonumber\\\left.
    +4\kappa_{n}(\kappa_{n}-\frac{2\pi r\rho 
    F'}{B})|X|^{2}\right\}.
\end{eqnarray}

\noindent Equation (\ref{separation}) implies that $\delta W(\psi )
<0$ for some $\psi$ is both a {\it necessary\/} and {\it sufficient\/}
condition for instability \citep{bernstein58}.  

Let us first address the question of stability of a plasma confined by
a stellar dipole magnetic field.  We approximate $F'$ by $\Delta
F/\psi_{a}$ where $\psi_{a}=a^{2}B_{0}/2$ and $\Delta F=\Delta
p_{0}/\rho_{0}$, with $\Delta p_{0}$ and $\rho_{0}$ being,
respectively, the overpressure and density at the top of the crust. 
Since $\kappa_{gn}=({\partial U}/{\partial n})/2c^{2}_{s0}\sim
-{\hat{b}_{r}}/{2h}$, $\kappa_{cn}\sim{\hat{b}_{r}}/{R}$,
$\hat{b}_{r}\sim r/R$ (for a dipole field) and the density scale height
$h=\cs2_{0}/g$ is much smaller than the neutron star radius $R$, the
second term on the RHS of Eq.~(\ref{crit}) is destabilizing, if
${\Delta}{\beta}\gtrsim a^{2}/hR$.

For the interchange modes ${\partial X}/{\partial l} =0$.  Because of
line tying, however, $X(0)=0$ with $l=0$ being chosen to correspond to
the top of the crust.  Expanding into Taylor series, we can
approximate $X\sim l$ for $l\lesssim h$.  From the structure of the
integral in Eq.~(\ref{crit}) we expect that the most unstable
displacement vanishes for $l\gg h$.

We shall now estimate the integral in Eq.~(\ref{crit}).  To lowest
order in $a/R$, where $a$ is the radius of the polar cap, $U =\cs2
l/h$.  We then find that $\delta W(\psi_{a})$ becomes negative only if

\begin{eqnarray}\label{crit2}
{\Delta}{\beta}\gtrsim\frac{R}{h}
\end{eqnarray}

\noindent where ${\Delta}{\beta}=8\pi\Delta p_{0}/B_{0}^{2}$.  The
above equation implies that for a low overpressure
(${\Delta}{\beta}\lesssim 1$), the plasma confined by the dipole
magnetic field of a neutron star is stable with respect to ballooning
modes.  Equation (\ref{crit2}) indicates that the plasma may become
unstable at high ${\Delta}\beta\gg 1$.  For such a high plasma
overpressure, however, the equilibrium magnetic field is distorted
away from the dipole field and has to be computed self-consistently. 
We consider this case in the next section.

\section{A simple high-$\Delta\beta$ equilibrium}\label{equilibrium}

In the limit $a\ll R$ the dipole magnetic field in the polar cap is
well approximated by a uniform field.  Let us therefore assume that
prior to the loading of the accretion column with plasma, the magnetic
field $\bB =B_{0}\hat{z}$.  We shall assume that at $z=0$ the flux at
the is unaffected by the accretion, i.e., $\psi (r,0)=r^{2}B_{0}/2$. 
The loading is assumed to occur on flux surfaces with
$\psi\leq\psi_{a}\equiv a^{2}B_{0}/2$.

Until now, the overpressure profile at the base of the accretion
region, which determines the function $F$, has not been specified.  As a
simple example we shall consider the following form of $F$,

\begin{equation}
        F(\psi )= F_{ext}+\Delta
        F\left[ 
        1-\left(\frac{\psi}{\psi_{a}}\right)^{2}\right]^{1+\varepsilon}
\end{equation}

\noindent for ${\psi}\leq {\psi_{a}}\equiv a^{2}B_{0}/2$, with
$\varepsilon\ll 1$; $F$ is a constant $F_{ext}$ for ${\psi}>{\psi_{a}}$. 
The gradient of $F$ is approximately linear for $\psi
<\psi_{a}(1-\varepsilon )$ and goes smoothly to zero at $\psi =\psi_{a}$. 
In the following we shall restrict our attention to the case of an
overpressure $\Delta p_{0}$ that is small compared to the total
pressure $p_{0}$, so that $\Delta F=\Delta p_{0}/\rho_{0}$.  In the
same approximation, the scale height $h$ is, to lowest order,
independent of $\psi$.

For such a pressure profile, the Grad-Shafranov equation
(Eq.~[\ref{eq5}]) inside the accretion column [$\psi
<\psi_{a}(1-\varepsilon )$] becomes approximately linear,

\begin{equation}\label{gs}
r\frac{\partial}{\partial r}\left(\frac{1}{r}\frac{\partial\psi}{\partial r}\right)
+\frac{\partial^{2}\psi}{\partial
z^{2}}=\frac{4{\Delta}\beta}{a^{2}}\left(\frac{r}{a}\right)^{2}\left(1-\alpha
\frac{z}{h}\right)^{\frac{1}{\alpha}}\psi.
\end{equation}

We anticipate that because ${\Delta}\beta\gg 1$ and $h\ll a$, the
radial excursion of the flux surfaces $\Delta r=\int dl\hatbr$
($\lesssim h$) will be small compared to $r$.  We can therefore
neglect the $r$-dependence on the RHS of Eq.~(\ref{gs}) and treat
$\chi\equiv r/a$ as a parameter.  With these approximations, we can
then solve Eq.~(\ref{gs}) by separation of variables,

\begin{equation}\label{flux}
   \psi (r,z)=\psi_{0}\chi^{2}f(\zeta)|_{\chi =r/a, \zeta =z/h},
\end{equation}

\noindent where $f$ satisfies the equation

\begin{equation}\label{gs2}
\frac{d^{2}f}{d \zeta^{2}}=\lambda^{2}\chi^{2}(1-\alpha\zeta
)^{\frac{1}{\alpha}}f
\end{equation}

\noindent ($\lambda\equiv 2h\sqrt{\Delta\beta}/a$).  We solve this
equation subject to the boundary conditions $f(0)=1$ and
$f'(\alpha^{-1})=0$.  The first of these conditions expresses
normalization; the second follows from the requirement that the field be
vertical in the region of vanishing overpressure.  By changing variables,
we can transform the above equation into a modified Bessel equation
which has the solution

\begin{equation}
f(\zeta)=u^{\nu}(\zeta )\left\{{\cal A}I_{-\nu}[u(\zeta )]+{\cal
B}I_{\nu}[u(\zeta )]\right\},
\end{equation}

\noindent where $u(\zeta )=2{\lambda\chi}(1-\alpha\zeta
)^{{\frac{1}{2\nu}}}/(1+2\alpha )$, $\nu =\alpha/(1+2\alpha )=(\gamma
-1)/(2\gamma -1)$, $I_{\pm\nu}$ are the modified Bessel functions, and
${\cal A}$ and ${\cal B}$ are the integration constants.  From the
condition $f'(\alpha^{-1})=0$ it follows that ${\cal B}=0$; the boundary
condition $f(0)=1$ then implies that

\begin{equation}\label{fluxz}
f(\zeta )=\left( 1-\alpha\zeta \right)^{{\frac{1}{2}}}
\frac{I_{-\nu}\left[\frac{2\lambda\chi}{1+2\alpha}\left( 1-\alpha\zeta
\right)^{{\frac{1}{2\nu}}}\right]}{I_{-\nu}\left(\frac{2\lambda\chi}{1+2\alpha}\right)}.
\end{equation}

\noindent An example of flux surfaces given by equations (\ref{flux})
and (\ref{fluxz}) is shown in Figure 2.  We see that even for
relatively large $\lambda$, the radial excursion $\Delta r$ of flux
surfaces is small compared to $r$, in accord with our assumptions.

\epsfig{file=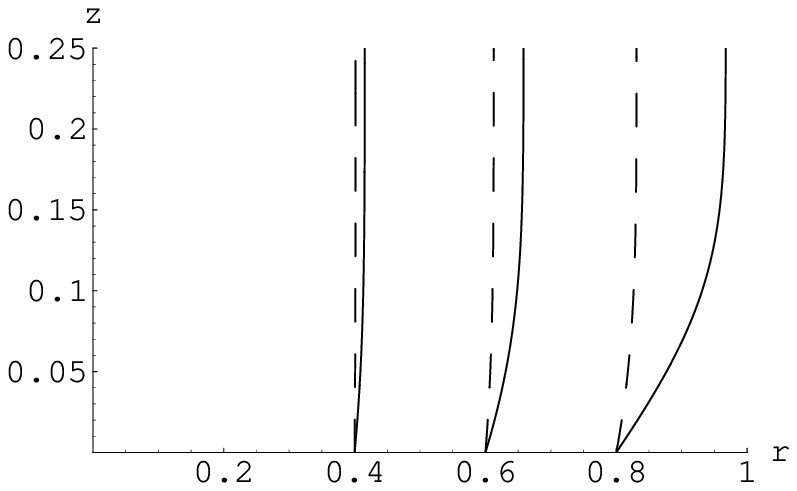,width=8.5cm}
\figcaption{\label{fig:flux} Poloidal cross-sections of flux 
surfaces for $\lambda =1$ (solid) and $\lambda =0.5$ (dashed), with 
$a/h=10$ ($r$ and $z$ are in units of $a$).}\bigskip

Let us now consider the case of $ a/h\lesssim{\Delta\beta}\ll
a^{2}/h^{2}$ which, as we shall see, is valid at the onset of
instability.  In this case, $\lambda\ll 1$ and we can expand the
Bessel functions into a power series.  In this approximation we find
that

\begin{equation}\label{approxflux}
   \psi (r,z)=\psi_{0}\frac{r^{2}}{a^{2}}\left[1 -
   \lambda^2\frac{r^{2}}{a^{2}}\frac{1 - \left(1 -
   \alpha\, {z}/{h}
   \right)^{2+\frac{1}{\alpha}}}{\left( 1 + \alpha\right)\left( 1 +
2\,\alpha \right)}
  \right].
\end{equation}

\noindent With the aid of the above expression we can now demonstrate
the self-consistency of our earlier assumptions.  We find that the
first term on the LHS of Eq.~(\ref{gs}) is smaller than the second one
as long as $\rho /\rho_{0}> (2h/a)^{2}$.  Thus for $h\ll a$ our
solution is indeed self-consistent in the region of plasma where
the most unstable displacement is expected to be localized (see
\S\ref{evaluation}).

From (\ref{approxflux}) the magnetic field components are
\begin{eqnarray}
B_{r}&\approx& 
\mu\left(\frac{\rho}{\rho_{0}}\right)^{\gamma}B_{0}\label{br}\\
B_{z}&\approx& B_{0}\label{bz}
\end{eqnarray}

\noindent where $\mu = 2\Delta\beta hx^{3}/\gamma a$.  From the above
it follows that the angle $\phi$ between the magnetic and 
gravitational fields is given by
\begin{eqnarray}\label{angle}
\tan\phi&\approx& \mu\left(1-\alpha\frac{z}{h}\right)^{\gamma/\alpha}.
\end{eqnarray}

\noindent From equation (\ref{angle}) one sees explicitly that the field line
distortion becomes large ($\tan\phi\gtrsim 1$) when $\mu\gtrsim 1$,
i.e., $\Delta\beta\gtrsim a/h$, as originally observed by
\citet{hameury83}.  This is, in fact, a simple consequence of
balancing overpressure $\Delta p/a$ with magnetic tension $\kappa
B^2/4\pi$ since $\kappa h\sim 1$ when the field distortion becomes
large.

\section{Criterion for the onset of instability}\label{evaluation}

We now use the equilibrium found in the previous section to evaluate the
energy integral.  Let us first change the integration variable in
Eq.~(\ref{crit}) from $l$ to $z$, such that $dl=dz/\hatbz$.  For the
nearly parabolic pressure profile considered earlier,

\begin{eqnarray}\label{crit3}
	\delta W(\psi )=\frac{m^{2}}{8\pi}\int\frac{dz}{r^{2}B_{z}}
	\left\{\hatbz^{2}\left|\frac{\partial X}{\partial z}\right|^{2}
	\right.\nonumber\\
	\left.
	+4\kappa_{n}\left(\kappa_{n}+\frac{\gamma\mu}{2h}\frac{B_{0}}{B}
	\frac{\rho}{\rho_{0}}\right)|X|^{2}\right\}.
\end{eqnarray}

\noindent In the approximation of Eq.~(\ref{approxflux}), we find that
to lowest order in $\lambda$

\begin{equation}\label{curvature}
\kappa_{n} \approx -(1+2\gamma\hatbz^{2})
\left(\frac{\rho_{0}}{\rho}\right)^{\alpha}\frac{\hatbr}{2h}.
\end{equation}

\noindent Changing the integration variable to $s=\tan\phi$, as given
by equation (\ref{angle}), we then find, with the aid of (\ref{br}),
(\ref{bz}), (\ref{angle}) and (\ref{curvature}),

\begin{eqnarray}\label{crit4}
	\left.\delta W(\psi )=\frac{\mu^{1-1/\gamma} m^{2}}{8\pi
	r^{2}\gamma h B_{0}}
	\int_{0}^{\mu}ds\frac{s^{1/\gamma}}{1+s^{2}}
	\right\{\gamma^{2}\left|\frac{\partial X}{\partial
	s}\right|^{2}\nonumber\\ \left.-(\gamma
	-1)\frac{(s^{2}-\mu_{0}^{2})(s^{2}+2\gamma +1)}{(s^{2}
	+1)^{2}}\left|X\right|^{2}\right\},
\end{eqnarray}

\noindent where $\mu_{0}=\sqrt{(\gamma+1)/(\gamma -1)}$.  We observe that the
integrand in equation (\ref{crit4}) is positive for $s\geq\mu_{0}$. 
Consequently, the most unstable displacement is expected to vanish for
$s\geq\mu_{0}$ which implies the instability is localized to the 
region where 

\begin{eqnarray}\label{localization}
\rho\geq\left(\frac{\mu_{0}}{\mu}\right)^{\frac{1}{\gamma}}\rho_{0}.
\end{eqnarray}

Since the line-tying in the crust requires that
$X(s=\mu )=0$, we therefore choose as a test function
$$
X(s)=X_{0}\sin\left(\pi\frac{s-\mu_{0}}{\mu-\mu_{0}}\right)
$$

For this test function and $\gamma = 4/3$, as appropriate for a plasma 
in the vicinity of the crust, we find that $\delta W(\psi )<0$ for 
$\mu>11.67$ for which it follows that

\begin{eqnarray}\label{betacrit}
\Delta{\beta}> 7.8\frac{a}{h}.
\end{eqnarray}

\noindent is a sufficient condition for instability.  For $\mu =11.67$,
equation (\ref{localization}) implies that the instability is
localized to within 0.93 scale height from the crust. 

The instability criterion (\ref{betacrit}) is obtained for the
equilibrium given by equation (\ref{approxflux}), derived in the
approximation of $\Delta\beta \ll (a/h)^{2}$.  Thus self-consistency
implies that the validity of (\ref{betacrit}) is limited to $h/a\ll
0.13$.  This condition is satisfied, albeit marginally, for the
parameters discussed in \S\ref{polar-caps}.  The instability threshold
for more general equilibria, with higher values of $h/a$, is at
present unknown.

\section{Mass of confined accreted matter}\label{mass}

Assuming that the crust is perfectly rigid, the amount of mass needed to
drive the instability is, from inequality~(\ref{betacrit}),
\begin{eqnarray}\label{eq:mass}
\Delta M &=& \pi a^2 \frac{B^2}{8\pi}\frac{\Delta\beta}{g}\nonumber \\
        &=&  0.73 B^2 a^3 \frac{\rho}{p}.
\end{eqnarray}
\noindent Substituting Eq. (\ref{scaleheight}) in Eq.~(\ref{eq:mass}), we
have
\begin{eqnarray}
   \Delta M &=& 3.8\ee{-13}\Msun
        \left(\frac{B}{10^{12}\gauss}\right)^2
        \left(\frac{a}{10^5\cm}\right)^3 \times \nonumber \\
&& \left(\frac{Z}{26}\right)^{2/3}\left(\frac{A}{56}\right)
        \left(\frac{5\ee{8}\K}{T}\right)
        \left(\frac{173}{\Gamma}\right). 
\end{eqnarray}
At a given accretion rate $\dot{M}$, the time necessary to accumulate
this mass is $\Delta M/\dot{M}=33\hour
(10^{-10}\Msun\yr^{-1}/\dot{M})$.  At present we cannot say what, if
any, are the observable effects of the instability, but the timescale
to accumulate a critical mass $\Delta M$ is well-suited for
monitoring.

\section{Discussion}\label{discussion}

Using the MHD energy principle, we have shown that the accreted
material on the polar cap of a strongly magnetized neutron star is
subject to the ballooning instability when the overpressure in the
vicinity of the crust exceeds the magnetic pressure by the factor $7.8
a/h$, where $a$ is the horizontal pressure gradient length scale
(identified with the polar cap radius) and $h$ is the density scale
height.  For typical polar cap parameters, the critical overpressure
is $\sim 10^{-4}$ of the total plasma pressure.  The instability is
localized to within a density scale height of the field line footpoint
(at the top of the crust) which implies that plasma is perturbed at
densities $(\rho/\rho_0) > 0.33$; for $\rho_0$ found from
equation~(\ref{eq:Gamma}) for a model atmosphere (cf.  Brown \&
Bildsten 1999), with a temperature of $7.6\ee{8}\K$, assuming a
locally Eddington accretion rate.  This implies
$\rho>10^{10}\GramPerCc$.  This density is higher than where the
accreted hydrogen and helium ignite which suggests that the ballooning
instability near the threshold for onset will not affect the nuclear
burning.  It is not clear, however, whether the instability can
degrade the confinement enough to prevent a further accumulation of
mass well above the instability threshold.

In our stability analysis we have omitted the contribution of the
Schwarzschild term to the MHD potential energy arguing that it is
small in highly degenerate plasmas of uniform composition immersed in
strong magnetic fields.  To test the validity of this approximation we
have performed a calculation along the same lines as in \S\ref{energy}
and \S\ref{ballooning}, including the buoyancy contribution, for
$\epsilon =4\times 10^{-4}$, for a realistic atmosphere (cf. Brown \&
Bildsten 1999).  We found that this inclusion increases the numerical
factor on the RHS of inequality (\ref{betacrit}) by 17\% for
$B=10^{12}$ G and by less than 1\% for $B\geq 5\times 10^{12}$.  Thus
the effect of buoyancy is indeed weak for magnetic fields of interest,
as long as $\epsilon$ is as small as assumed.

The latter assumption, as well as that of composition uniformity, is
violated in the electron-capture layers.  The location of these layers
is not precisely known.  \citet{haensel90} found that electron capture
layers closest to the top of the crust occur at densities $10^{10}$ g
cm$^{-3}$ and $8\times10^{10}$ g cm$^{-3}$.  Since the instability
found in the present paper is localized within a density scale height
from the crust which for the model atmosphere implies localization
between densities $10^{10}$ and $3\times10^{10}$ g cm$^{-3}$, we
expect that the electron capture effects would not significantly
modify our conclusions.

In our analysis the crust was modelled as a rigid boundary in which
the magnetic field is anchored.  For this approximation to be
justified the lateral distortion of the crust in response to the
overpressure must be much smaller than the magnetic field line
distortion in the overlying fluid.  This is true when
$\Delta\beta/\beta$ exceeds $\alpha_{s}\Delta a/a$ where $\Delta a\sim
(h^{2}/a)\Delta\beta$ is the displacement of flux surfaces in the
ocean (cf.  \S\ref{equilibrium}), i.e., when $\beta >
(a/h)^{2}/\alpha_{s}$; here $\alpha_{s}p$ is the crust shear modulus. 
Estimates \citep{pandharipande.pines.ea:neutron,strohmayer..ea:shear}
find $\alpha_{s}\sim 10^{-3}-10^{-2}$, which for the characteristic
density scale height and polar cap radius (cf.  \S\ref{intro}) implies
that $\beta\gtrsim 10^{4}-10^{5}$.  For the model atmosphere, this
condition is satisfied for $B\lesssim 5\times 10^{12}$ G.

The imperfect rigidity of the crust may affect the estimate of the
accreted mass necessary to destabilize the ballooning mode
(eq.~[\ref{eq:mass}]).  \citet{ushomirsky.cutler.ea:deformations}
found that the crust tended to sink in response to a lateral density
discontinuity.  While their findings are not directly applicable to
our case, it is nevertheless conceivable that the crust could deform
sufficiently to significantly reduce the applied overpressure.  The
mass required to drive the instability would then be higher than our
estimate.

The effect of the instability is an enhanced cross-field transport of
matter and a resulting decrease in the plasma confinement time in the
polar cap.  To find the magnitude of this anomalous transport requires
analysis of nonlinear effects which are beyond the scope of this
paper.  The analysis of these effects and, in particular, the question
of the existence of a steady state, in which the transport is
sufficiently rapid to balance the influx of the accreted matter, are
relegated to future publications.

\acknowledgements The authors thank Ellen Zweibel for pointing out the
potential importance of the stabilizing effect of the Schwarzschild
term in equation (\ref{energy-1}).  Helpful discussions with Thierry
Emonet, Timur Linde, and Dmitri Uzdensky are also acknowledged.  This
research was supported by the ASCI Center for Astrophysical
Thermonuclear Flashes at the University of Chicago under Department of
Energy contract B341495.  EFB acknowledges support from an Enrico
Fermi Fellowship at the University of Chicago.

\end{document}